# Reliable operation in high-mobility indium oxide thin film transistors

Prashant R. Ghediya, Yusaku Magari*, Hikaru Sadahira, Takashi Endo, Mamoru Furuta, Yuqiao Zhang, Yasutaka Matsuo & Hiromichi Ohta*

P. R. Ghediya, Y. Magari, T. Endo, Y. Matsuo, H. Ohta
Research Institute for Electronic Science, Hokkaido University, N20W10, Kita, Sapporo 001-0020, Japan
E-mail: yusaku.magari@es.hokudai.ac.jp, hiromichi.ohta@es.hokudai.ac.jp

H. Sadahira
Graduate School of Information Science and Technology, Hokkaido University, N20W10, Kita, Sapporo 001-0020, Japan

M. Furuta
School of Environmental Science and Engineering, Kochi University of Technology, Kami, Kochi, 782-8502, Japan

Y. Zhang
Institute of Quantum and Sustainable Technology, Jiangsu University, Zhenjiang 212013, China

**Abstract**

Transparent oxide semiconductors (TOSs) based thin-film transistors (TFTs) that exhibit higher field effect mobility ($\mu_{FE}$) are highly required toward the realization of next-generation displays. Among numerous types of TOS-TFTs, $In_2O_3$-based TFTs are the front-running candidate because they exhibit the highest $\mu_{FE}$ ~100 cm$^2$ V$^{-1}$ s$^{-1}$. However, the device operation of $In_2O_3$ TFTs is unreliable; a large voltage shift occurs especially when negative gate bias is applied due to adsorption/desorption of gas molecules. Although passivation of the TFTs is used to overcome such instability, previously proposed passivation materials did not improve the reliability. Here, we show that the $In_2O_3$ TFTs passivated with $Y_2O_3$ and $Er_2O_3$ films are highly reliable and do not show threshold voltage shifts when applying gate bias. We applied positive and negative gate bias to the $In_2O_3$ TFTs passivated with various insulating oxides and found that only the $In_2O_3$ TFTs passivated with $Y_2O_3$ and $Er_2O_3$ films did not exhibit threshold voltage shifts. We observed that only the $Y_2O_3$ grew heteroepitaxially on the $In_2O_3$ crystal. This would be the origin of the high reliability of the $In_2O_3$ TFTs passivated with $Y_2O_3$ and $Er_2O_3$ films. This finding accelerates the development of next-generation displays using high-mobility $In_2O_3$ TFTs.

## Introduction

Transparent oxide semiconductor (TOS)-based thin-film transistors (TFTs) have been widely used as the backplane of flat panel displays (FPDs), such as liquid crystal displays (LCDs) and organic light-emitting diodes (OLEDs)[1-4]. Their performance is characterized by the field effect mobility ($\mu_{FE}$). The $\mu_{FE}$ of amorphous $InGaZnO_4$ (a-IGZO) TFTs[5,6], which has been commercialized, is ~10 $cm^2\ V^{-1}\ s^{-1}$. However, even higher-definition (*e.g.,* super high definition, SHD) FPDs that are operatable at higher frequencies (>240 Hz) demand $\mu_{FE}$ ~100 $cm^2\ V^{-1}\ s^{-1}$, one order of magnitude higher than that of a-IGZO TFTs[7-10]. Therefore, numerous efforts have been made thus far to find good TOSs that show high $\mu_{FE}$.[11-17] Among numerous reports on TOS-TFTs, Magari *et al*. reported that hydrogen incorporated $In_2O_3$ TFTs stand out high $\mu_{FE}$ ~140 $cm^2\ V^{-1}\ s^{-1}$ [17]. Although the influence of the grain boundary scattering of electrons in polycrystalline materials is a major concern, the low-temperature (~300 °C) solid-phase crystallization technique is effective in increasing the lateral grain size and mobility and reducing the subgap defects in $In_2O_3$ films. Thus, $In_2O_3$ TFTs have been front-running candidates.

However, there is a serious drawback to the operation of $In_2O_3$ TFTs, which makes them inappropriate for practical applications. In real applications, the TFT operation must be stable against gate bias applications. Current $In_2O_3$ TFTs exhibit a large negative threshold voltage ($V_{th}$) shift when a negative gate bias stress is applied (**Fig. 1a**). In the case of a-IGZO TFTs, the origin of the $V_{th}$ shift is mainly gas adsorption/desorption, and such a $V_{th}$ shift is suppressed by covering the active material layer (passivation).[18,19] As a passivation material, a-$SiO_2$ is widely used, and the $V_{th}$ shift of a-IGZO TFTs is successfully suppressed.[20-24] Therefore, Magari *et al.*[17] deposited a-$SiO_2$ film as the passivation layer of the $In_2O_3$ TFTs, but the resultant TFTs showed a large $V_{th}$ shift (~5 V) after applying negative gate bias stress. Thus, a-$SiO_2$ passivation is useless in improving the reliability of $In_2O_3$ TFTs.

To overcome this difficulty and realize ideal $In_2O_3$ TFTs with reliability (**Fig. 1b**), we passivated the $In_2O_3$ active layer surface of $In_2O_3$ TFTs with various insulating oxides, as schematically shown in **Fig. 1c**, including $HfO_2$, $Al_2O_3$, $Y_2O_3$, $Er_2O_3$, $Gd_2O_3$, $Yb_2O_3$, $Sm_2O_3$, and $Nd_2O_3$. We selected amorphous $HfO_2$ and $Al_2O_3$ films because they are often used as the passivation layers for TOS TFTs.[25-27] Further, we selected $Ln_2O_3$ (*Ln*: Lantanoid, *Ln* = Y, Er, Gd, Yb, Sm, and Nd) because the ionic radius of the $Ln^{3+}$ is close to that of $In^{3+}$ and the crystal structure of $Ln_2O_3$ (C-rare earth structure: *Ln* = Y, Er, Gd, and Yb, B-rare earth structure: *Ln* = Sm, A-rare earth structure: *Ln* = Nd) is similar to that of $In_2O_3$ (C-rare earth structure). Note that Nomura *et al*. reported that $Y_2O_3$ passivation effectively improves the reliability of a-IGZO TFTs.[28] We expected that heteroepitaxial growth of $Ln_2O_3$ would occur on $In_2O_3$[29] and the reliability would be improved. Interestingly, only the $In_2O_3$ TFTs passivated with $Y_2O_3$ and $Er_2O_3$ films showed excellent reliability, whereas those passivated with all other insulating oxides exhibited poor reliability.

Here, we show that $In_2O_3$-TFTs passivated with $Y_2O_3$ and $Er_2O_3$ films are highly reliable; the $V_{th}$ shifts are negligible after applying a gate bias. We applied positive and negative gate biases to $In_2O_3$ TFTs passivated with various insulating oxides and found that only the $In_2O_3$ TFTs passivated with $Y_2O_3$ and $Er_2O_3$ films did not exhibit $V_{th}$ shifts. We observed that only $Y_2O_3$ grew heteroepitaxially on the $In_2O_3$ crystals. This could be the origin of the high reliability of the $In_2O_3$ TFTs passivated with $Y_2O_3$ and $Er_2O_3$. The present findings are expected to accelerate the development of next-generation displays using $In_2O_3$ TFTs showing high $\mu_{FE}$.

## Results and discussion

We fabricated bottom-gate top-contact $In_2O_3$ TFTs on ITO-coated alkali-free glass substrates. The ITO was used as the bottom gate electrode. We used a 100-nm-thick $AlO_x$ film as a gate insulator. The dielectric permittivity ($\varepsilon_r$) and capacitance ($C_i$) of the $AlO_x$ films were 8 and 70 nF $cm^{-2}$, respectively. We deposited a 5-nm-thick amorphous $In_2O_3$ film on the $AlO_x$ film at room temperature through a stencil mask. The multilayer film was heated at 300 °C in air for 0.5 h to crystalize the $In_2O_3$ active layer. Then, 100-nm-thick indium tin oxide (ITO) source/drain electrodes were deposited through a stencil mask. The channel length ($L$) and width ($W$) were 200 and 400 µm, respectively. After that, ~50-nm-thick $HfO_2$, $Al_2O_3$, and $Ln_2O_3$ (*Ln*: Lanthanoid = Y, Er, Gd, Yb, Sm, and Nd) films were deposited as the passivation layer through a stencil mask without substrate heating. Finally, the TFT was thermally annealed at 370 °C in air for 0.5 h. After the thermal annealing, $Ln_2O_3$ films were polycrystalline whereas $HfO_2$ and $Al_2O_3$ films remained amorphous (**Fig. S1**).

The transistor characteristics were measured using a semiconductor device analyzer (B1500A, Agirent) by applying a constant drain voltage ($V_d$) of 5 V. We measured the transfer characteristics after applying a positive gate bias stress (PBS; $V_g$ = +20 V, electric field $E$ = 2 MV cm$^{-1}$) and a negative gate bias stress (NBS; $V_g$ = −20 V, electric field $E$ = −2 MV cm$^{-1}$) for up to 5000 s to the TFTs at room temperature in air. Details of our TFT fabrication and measurement are described in the **Method** section.

First, we measured the transistor characteristics of the $In_2O_3$ TFTs without a passivation layer (**Fig. 2**). The TFT showed excellent transistor characteristics before applying bias stress (Holding time = 0 s in **Fig. 2a**); the field effect mobility ($\mu_{FE}$) of ~85 cm$^2$ V$^{-1}$ s$^{-1}$, threshold voltage ($V_{th}$) of −3.2 V, subthreshold swing of ~0.22 V decade$^{-1}$, and TFT did not show a hysteresis behavior. However, after applying PBS for 5000 s, a positive $V_{th}$ shift of +1 V was observed (**Fig. 2a**). After applying NBS for 5000 s, the negative $V_{th}$ shift of −10 V occurred (**Fig. 2b**). These results demonstrate that the polycrystalline $In_2O_3$ TFTs without a passivation layer are unreliable.

Then, we measured the transistor characteristics of the $In_2O_3$ TFTs with passivation layers (**Figs. S2−S9**). Although $HfO_2$ and $Al_2O_3$ play as good passivation materials for a-IGZO TFTs, the $In_2O_3$ TFTs passivated with $HfO_2$ (**Fig. S2**) and $Al_2O_3$ (**Fig. S3**) show a large $V_{th}$ shift especially after NBS ($HfO_2$: −4.4 V, $Al_2O_3$: −7.0 V), showing unreliability of the TFTs. Importantly, the $In_2O_3$ TFTs passivated with the $Y_2O_3$ (**Fig. S4**) and $Er_2O_3$ (**Fig. S5**) films exhibited excellent reliability, and the $V_{th}$ shifts were negligible after both PBS and NBS. The TFTs passivated with the $Gd_2O_3$ (**Fig. S6**) and $Yb_2O_3$ (**Fig. S7**) films showed good reliability. However, the TFTs passivated with the $Sm_2O_3$ (**Fig. S8**) and $Nd_2O_3$ (**Fig. S9**) films exhibited a large $V_{th}$ shift after both PBS and NBS. The $V_{th}$ shift results after 5000 s application of PBS and NBS are summalized in **Figs. 2c** and **2d**, respectively. From these results, we conclude that the $Y_2O_3$ and $Er_2O_3$ films are excellent passivation materials for realizing reliable $In_2O_3$ TFTs.

We also realized that the performance degradation of the $In_2O_3$ TFTs did not occur using $Y_2O_3$ and $Er_2O_3$ film passivation. **Figure 3** shows the bias stress time dependence of the $\mu_{FE}$ (**Figs. 3a** and **3b**) and subthreshold swing (**Figs. 3c** and **3d**) under PBS and NBS. The $\mu_{FE}$ and subthreshold swing values of the $In_2O_3$ TFT without passivation drastically changed with stress time, showing unreliability. In contrast, the $\mu_{FE}$ and subthreshold swing values of the $In_2O_3$ TFTs passivated with $Y_2O_3$ and $Er_2O_3$ are highly stable against the bias stress application. The $\mu_{FE}$ of the $Y_2O_3$ passivated TFT was ~78 $cm^2$ $V^{-1}$ $s^{-1}$ and that of the $Er_2O_3$ passivated TFT was ~70 $cm^2$ $V^{-1}$ $s^{-1}$. The subthreshold swing of both TFTs was ~0.1 V $decade^{-1}$. These results demonstrate that a reliable device operation in $In_2O_3$ TFTs can be achieved using $Y_2O_3$ and $Er_2O_3$ films as passivation layers.

Here, we discuss the origin of the reliability of $In_2O_3$ TFT passivated with $Y_2O_3$ and $Er_2O_3$ films. We hypothesized that the $Y_2O_3$ and $Er_2O_3$ films grew heteroepitaxially on the $In_2O_3$ films because these oxides have the same crystal structure as $In_2O_3$ (C-rare-earth structure). To confirm this, we prepared ~100-nm-thick rare-earth oxide ($Ln_2O_3$) ($Ln$ = Y, Nd, and Sm) films on (111)-oriented 10%-Sn-doped $In_2O_3$ epitaxial films using the same procedure as for passivation layer fabrication. **Figure 4** shows cross-sectional high-angle annular dark-field (HAADF) scanning transmission electron microscopy (STEM) images of the $Ln_2O_3$ films. Although several strong contrasts, indicated by yellow arrows owing to cracks, were observed in the $Sm_2O_3$ and $Nd_2O_3$ films, such microstructures were not observed in the $Y_2O_3$ film. Weak contrasts were observed around the heterointerfaces between $Sm_2O_3/In_2O_3$ and $Nd_2O_3/In_2O_3$. Interestingly, the lattice structures of $Y_2O_3$ are visible together with that of $In_2O_3$ in the magnified HAADF-STEM images (**Figs. 5a** and **5b**), indicating that heteroepitaxial growth occurred. In contrast, the magnified images of the heterointerfaces of the $Sm_2O_3/In_2O_3$ (**Fig. 5c**) and $Nd_2O_3/In_2O_3$ (**Fig. 5d**) visualize the existence of an amorphous structure, which is the origin of the contrast around the heterointerfaces. These results reveal that heteroepitaxial growth of the passivation materials is the key to improving the reliability of $In_2O_3$ TFTs.

To further clarify the origin that heteroepitaxial growth of $Y_2O_3$ occurred on $In_2O_3$, we plotted the relationship between the ionic radius and metal−oxygen bond ($M$−O) length in various $Ln_2O_3$ (**Fig. 6a**). $Ln_2O_3$ has three different crystal structures (*i.e.* C-, B-, and A-type) depending on the ionic radius. The C-type structures ($Yb_2O_3$, $Er_2O_3$, $Y_2O_3$, and $Gd_2O_3$) are cubic bixbyite and have the same crystal structure as $In_2O_3$ with an ionic radius of 0.80 Å. As shown in light blue in **Fig. 6a**, the $M$−O lengths of Er−O and Y−O are close to In−O. Therefore heteroepitxial growth of $Y_2O_3$ occurred on $In_2O_3$. We also plotted the $V_{th}$ shifts after PBS and NBS of the $In_2O_3$ TFTs as a function of the difference in $M$−O length from In−O length ($\Delta l_{(M-O)-(In-O)}$) (**Fig. 6b**). The $V_{th}$ shifts are significant for both PBS and NBS when the shortest $\Delta l_{(M-O)-(In-O)}$ of the passivation layer exceeds 0.15 Å ($Gd_2O_3$). In contrast, the $V_{th}$ shifts are negligible when the shortest $\Delta l_{(M-O)-(In-O)}$ is below 0.10 Å ($Y_2O_3$ and $Er_2O_3$). These results indicate that the bias stress stability is highly dependent on the $\Delta l_{(M-O)-(In-O)}$ of the passivation layer.

Finally, we like to discuss the reason why the heteroepitaxial growth of the passivation material effectively improves the bias stress stability of $In_2O_3$ TFTs. Generally, gas molecules ($O_2$, $H_2O$, $CO_2$, etc.) are adsorbed on the surface of oxide semiconductors such as IGZO[18,19], $In_2O_3$[30], and $SnO_2$[31,32] and drastically suppress the conductivity of the oxide semiconductors. We performed thermal desorption spectroscopy (TDS) measurements of the $In_2O_3$ films to determine the absorbed gas molecules (**Fig. S10**). We confirmed that the $H_2$ and CO desorption are negligible while small desorption of $O_2$ and large desorption of $H_2O$ occur from $In_2O_3$ film at relatively low temperatures (~60 °C). These results suggest that the desorption of the $OH^-$ ion occurs during the NBS test of the $In_2O_3$ TFTs and the ion leaves carrier electrons in the $In_2O_3$ film. Thus, NBS increases the residual carrier electron concentration, resulting in significant $V_{th}$ shifts toward the negative side after NBS. Consequently, the heteroepitaxially grown $Y_2O_3$ and $Er_2O_3$ passivation layers would effectively prevent the desorption of $OH^-$ ions from the $In_2O_3$ films. These findings are expected to accelerate the development of next-generation FPDs using $In_2O_3$ TFTs.

**Summary**

We demonstrated a key interface engineering strategy to stabilize the device operation of high-mobility $In_2O_3$ TFTs by applying $Y_2O_3$ and $Er_2O_3$ films as a passivation layer. We applied positive and negative gate bias to the $In_2O_3$ TFTs passivated with various insulating oxides and found that only the $In_2O_3$ TFTs passivated with $Y_2O_3$ and $Er_2O_3$ films did not exhibit degradations in $V_{th}$, $\mu_{FE}$, and subthreshold swing. We observed that only the $Y_2O_3$ films grew heteroepitaxially on the $In_2O_3$ crystals. Consequently, protecting the $In_2O_3$ back-channel surface with $Ln_2O_3$ films, featuring similar crystal structure and $M$−O length as $In_2O_3$, prevents the desorption/absorption of gas molecules. This could be the origin of the high reliability of the $In_2O_3$ TFTs passivated with $Y_2O_3$ and $Er_2O_3$. These findings are expected to accelerate the development of next-generation displays using high-mobility $In_2O_3$ TFTs.

**Methods**

## Fabrication of $In_2O_3$ TFTs

Bottom-gate top-contact $In_2O_3$ TFTs were fabricated on ITO-coated alkali-free glass substrates (EAGLE XG®, Corning®). First, a 100-nm-thick $AlO_x$ gate insulator was deposited on the substrate by atomic layer deposition. The dielectric permittivity ($\varepsilon_r$) and capacitance ($C_i$) of the $AlO_x$ films were 8 and 70 nF cm$^{-2}$, respectively. Subsequently, a 5-nm-thick $In_2O_3$ film was deposited on the $AlO_x$ film via pulsed laser deposition (PLD) under an oxygen pressure of 3 Pa at room temperature through a stencil mask. The multilayer film was annealed at 300 °C in ambient air for 30 min. Subsequently, 100-nm-thick indium tin oxide (ITO) source/drain electrodes were deposited by PLD through a stencil mask. The channel length ($L$) and width ($W$) were 200 and 400 µm, respectively. Subsequently, ~50-nm-thick $HfO_2$, $Al_2O_3$, and rare-earth oxide ($Ln_2O_3$) ($Ln$: Lanthanoid = Y, Er, Gd, Yb, Sm, and Nd) films were deposited via PLD through a stencil mask without substrate heating. These films served as passivation layers. Finally, the TFT was thermally annealed at 370 °C in air for 30 min.

## Reliability tests of the $In_2O_3$ TFTs

The transfer characteristics of the $In_2O_3$ TFTs were measured in the dark using a semiconductor device analyzer (B1500A, Agilent). Positive gate bias stress (PBS, $V_g$ = +20 V) and negative gate bias stress (NBS, $V_g$ = −20 V) tests were conducted for $In_2O_3$ TFTs for 5000 s at room temperature in air ambient. The field effect mobility ($\mu_{FE}$) was calculated from the linear transfer characteristics at a drain voltage ($V_d$) of 5 V using equation (1).

$$\mu_{FE} = \frac{g_m}{C_i \frac{W}{L} V_d}, \qquad (1)$$

where $g_m$ is the transconductance, $C_i$ is the oxide capacitance of the gate insulator (70 nF cm$^{-2}$ for $AlO_x$), and $V_d$ is the drain voltage. The subthreshold swing was extracted from $V_g$, which required an increase in the drain current ($I_d$) from 1 to 10 pA.

## Microstructure analyses of the $Ln_2O_3$ films

To visualize detailed microstructures, including lattice images, single-crystalline $In_2O_3$ substrates are necessary. First, we prepared atomically flat 10%-Sn-doped $In_2O_3$ epitaxial films on (111)-oriented YSZ single crystal substrates by PLD. Details of the 10%-Sn-doped $In_2O_3$ epitaxial film growth have been published elsewhere[29]. We then deposited $Ln_2O_3$ ($Ln$ = Y, Sm, and Nd) films on the $In_2O_3$ epitaxial films by PLD under the same conditions used for TFT fabrication. Finally, the bilayer films were thermally annealed at 370 °C in air for 30 min. The cross-sectional microstructures of the $Ln_2O_3$ films were observed using a high-angle annular dark-field (HAADF) scanning transmission electron microscope (STEM; JEM-ARM 200CF, JEOL Co. Ltd.) operated at 200 keV. The incident e-beam direction was $\langle 1\bar{1}0 \rangle$ $In_2O_3$.

### Thermal desorption spectroscopy of the $In_2O_3$ films

The desorption gases ($H_2$, $H_2O$, CO, and $O_2$) from the $In_2O_3$ film were measured by thermal desorption spectroscopy (TDS, TDS1200II, ESCO) with a raising temperature rate of 60 °C $min^{-1}$. The base pressure of the system was $5 \times 10^{-7}$ Pa. Si substrates with 100-nm-thick thermally grown $SiO_2$ were used for the TDS measurements.

**Supplementary Information**

Supplementary Figs. S1−S10 and Table S1

**Acknowledgements**

P.R.G. and Y. Magari contributed equally to this work. The authors thank Yuko Mori and Naomi Hirai for the STEM observations. This study was supported by Grant-in-Aid for Innovative Areas (19H05791) from the Japan Society for the Promotion of Science (JSPS). Y. Magari was supported by a Grant-in-Aid from the JSPS (22K14303). M.F. was supported by a Grant-in-Aid from JSPS (22K04200). Y.Z. was supported by the National Natural Science Foundation of China (Grant No. 52202242), Ministry of Science and Technology of the PRC (G2022014136L), National Science Foundation of the Jiangsu Higher Education Institutions of China (22KJB430002), and Start-Up Fund of Jiangsu University (5501310015). H.O. was supported by a Grant-in-Aid for Scientific Research A (22H00253) from the JSPS. Part of this work was supported by the Advanced Research Infrastructure for Materials and Nanotechnology in Japan (JPMXP1223HK0082) of the Ministry of Education, Culture, Sports, Science, and Technology (MEXT). Part of this work was supported by the Crossover Alliance to Create the Future with People, Intelligence, and Materials, and by the Network Joint Research Center for Materials and Devices.

**References**

1 Kamiya, T. & Hosono, H. Material Characteristics and Applications of Transparent Amorphous Oxide Semiconductors. *NPG Asia Mater.* **2**, 15-22 (2010).

2 Fortunato, E., Barquinha, P. & Martins, R. Oxide Semiconductor Thin-Film Transistors: A Review of Recent Advances. *Adv. Mater.* **24**, 2945-2986 (2012).

3 Hosono, H. How we made the IGZO transistor. *Nat. Electron.* **1**, 428-428 (2018).

4 *Amorphous Oxide Semiconductors: IGZO and Related Materials for Display and Memory*. (Wiley, 2022).

5 Nomura, K. *et al.* Room-temperature fabrication of transparent flexible thin-film transistors using amorphous oxide semiconductors. *Nature* **432**, 488-492 (2004).

6 Kim, Y. H. *et al.* Flexible metal-oxide devices made by room-temperature photochemical activation of sol-gel films. *Nature* **489**, 128-U191 (2012).

7 Kwon, J. Y. & Jeong, J. K. Recent progress in high performance and reliable n-type transition metal oxide-based thin film transistors. *Semicond. Sci. Technol.* **30**, 024002 (2015).

8 Hara, Y. *et al.* IGZO-TFT technology for large-screen 8K display. *J. Soc. Inf. Display* **26**, 169-177 (2018).

9 Hendy, I., Brewer, J. & Muir, S. Development of High-Performance IGZO Backplanes for Displays. *Information Display* **38**, 60-68 (2022).

10 Kim, T. *et al.* Progress, Challenges, and Opportunities in Oxide Semiconductor Devices: A Key Building Block for Applications Ranging from Display Backplanes to 3D Integrated Semiconductor Chips. *Adv. Mater.* **35**, e2204663 (2023).

11 Dehuff, N. L., Kettenring, E. S., Hong, D., Chiang, H. Q. & Wager, J. F. Transparent thin-film transistors with zinc indium oxide channel layer. *J. Appl. Phys.* **97**, 064505 (2005).

12 Cho, S. H. *et al.* Highly Stable, High Mobility Al:SnZnInO Back-Channel Etch Thin-Film Transistor Fabricated Using PAN-Based Wet Etchant for Source and Drain Patterning. *IEEE Trans. Electron Dev.* **62**, 3653-3657 (2015).

13 Sheng, J. *et al.* Amorphous IGZO TFT with High Mobility of ~70 $cm^2/Vs$ via Vertical Dimension Control Using PEALD. *ACS Appl. Mater. Interfaces* **11**, 40300-40309 (2019).

14 Shiah, Y. S. *et al.* Mobility-stability trade-off in oxide thin-film transistors. *Nat. Electron.* **4**, 800-807 (2021).

15 Lee, J. *et al.* Hydrogen-Doping-Enabled Boosting of the Carrier Mobility and Stability in Amorphous IGZTO Transistors. *ACS Appl. Mater. Interfaces* **14**, 57016-57027 (2022).

16 Rabbi, M. H. *et al.* Polycrystalline InGaO Thin-Film Transistors with Coplanar Structure Exhibiting Average Mobility of approximately 78 $cm^2 V^{-1} s^{-1}$ and Excellent Stability for Replacing Current Poly-Si Thin-Film Transistors for Organic Light-Emitting Diode Displays. *Small Methods* **6**, e2200668 (2022).
17 Magari, Y., Kataoka, T., Yeh, W. C. & Furuta, M. High-mobility hydrogenated polycrystalline $In_2O_3$ ($In_2O_3$:H) thin-film transistors. *Nat. Commun.* **13**, 1078 (2022).
18 Jeong, J. K., Yang, H. W., Jeong, J. H., Mo, Y. G. & Kim, H. D. Origin of threshold voltage instability in indium-gallium-zinc oxide thin film transistors. *Appl. Phys. Lett.* **93**, 123508 (2008).
19 Conley, J. F. Instabilities in Amorphous Oxide Semiconductor Thin-Film Transistors. *IEEE Trans. Dev. Mater. Reliability* **10**, 460-475 (2010).
20 Chen, T. C. *et al.* Light-induced instability of an InGaZnO thin film transistor with and without $SiO_x$ passivation layer formed by plasma-enhanced-chemical-vapor-deposition. *Appl. Phys. Lett.* **97**, 192103 (2010).
21 Chowdhury, M. D. H. *et al.* Effect of $SiO_2$ and $SiO_2/SiN_x$ Passivation on the Stability of Amorphous Indium-Gallium Zinc-Oxide Thin-Film Transistors Under High Humidity. *IEEE Trans. Electron Dev.* **62**, 869-874 (2015).
22 Zheng, L. L. *et al.* High-Performance Unannealed a-InGaZnO TFT With an Atomic-Layer-Deposited $SiO_2$ Insulator. *IEEE Electron. Dev. Lett.* **37**, 743-746 (2016).
23 Aman, S. G. M., Koretomo, D., Magari, Y. & Furuta, M. Influence of Deposition Temperature and Source Gas in PE-CVD for $SiO_2$ Passivation on Performance and Reliability of In-Ga-Zn-O Thin-Film Transistors. *IEEE Trans. Electron Dev.* **65**, 3257-3263 (2018).
24 Xiao, X. *et al.* Room-Temperature-Processed Flexible Amorphous InGaZnO Thin Film Transistor. *ACS Appl. Mater. Interfaces* **10**, 25850-25857 (2018).
25 Yang, S. *et al.* Improvement in the photon-induced bias stability of Al-Sn-Zn-In-O thin film transistors by adopting $AlO_x$ passivation layer. *Appl. Phys. Lett.* **96**, 213511 (2010).
26 Ko, Y. *et al.* The effects of a $HfO_2$ buffer layer on $Al_2O_3$-passivated indium-gallium-zinc-oxide thin film transistors. *Phys. Status Solidi. RRL* **5**, 403-405 (2011).
27 Nomura, K., Kamiya, T. & Hosono, H. Stability and high-frequency operation of amorphous In-Ga-Zn-O thin-film transistors with various passivation layers. *Thin Solid Films* **520**, 3778-3782 (2012).

28 Nomura, K., Kamiya, T. & Hosono, H. Highly stable amorphous In-Ga-Zn-O thin-film transistors produced by eliminating deep subgap defects. *Appl. Phys. Lett.* **99**, 053505 (2011).

29 Ohta, H. *et al.* Transparent organic thin-film transistor with a laterally grown non-planar phthalocyanine channel. *Adv. Mater.* **16**, 312 (2004).

30 Gurlo, A. *et al.* Grain size control in nanocrystalline $In_2O_3$ semiconductor gas sensors. *Sensor Actuat. B* **44**, 327-333 (1997).

31 Yamazoe, N. New Approaches for Improving Semiconductor Gas Sensors. *Sensor Actuat B-Chem* **5**, 7-19 (1991).

32 Liang, D. D., Zhang, Y. Q., Hai, J. C. & Ohta, H. Electric field thermopower modulation analyses of the operation mechanism of transparent amorphous $SnO_2$ thin-film transistor. *Appl. Phys. Lett.* **116**, 143503 (2020).

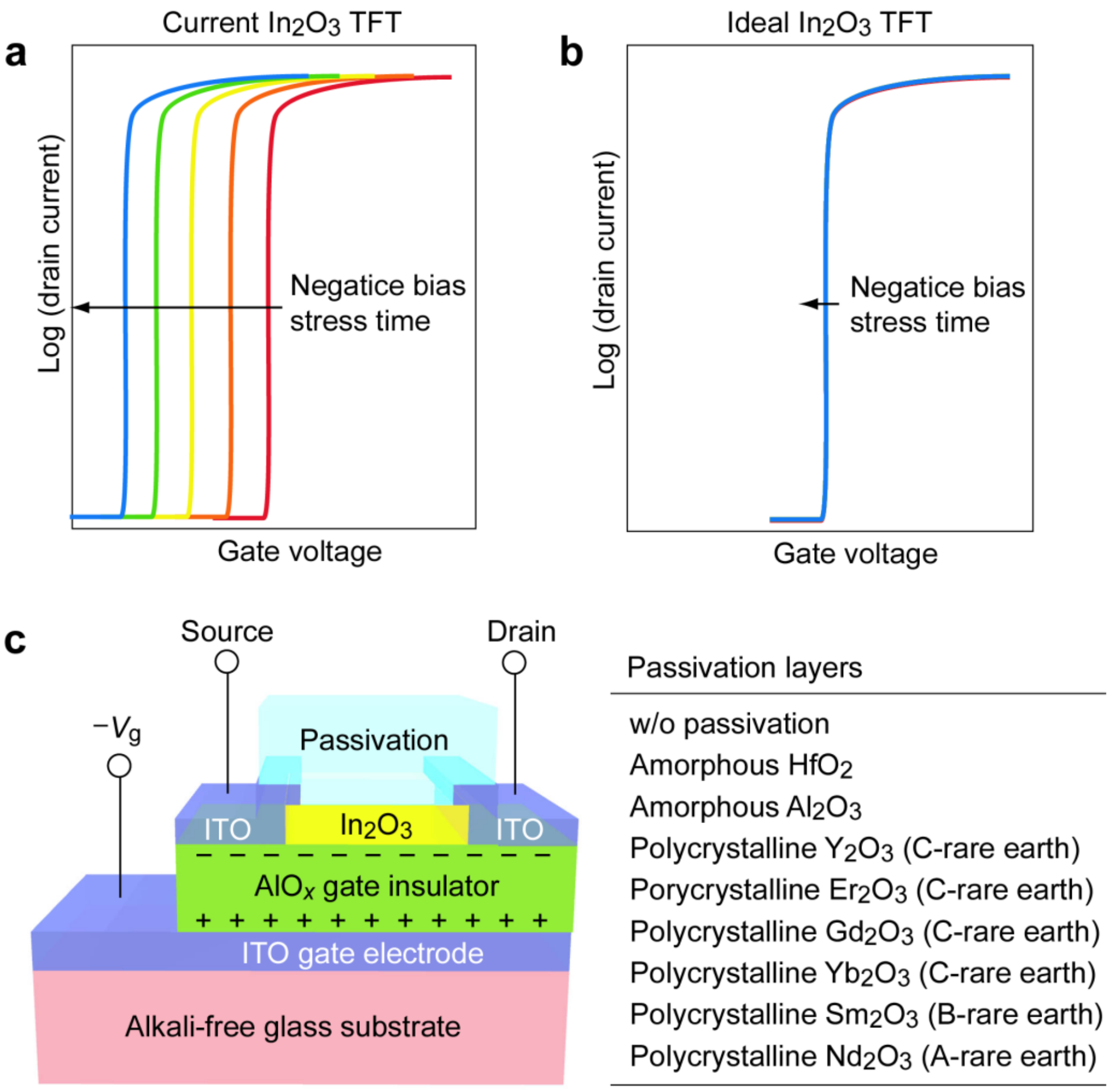

**Figure 1: Passivation strategy towards the improvement of reliability of high-mobility $In_2O_3$ TFTs.**

Schematic transistor characteristics after applying bias stress of (a) current and (b) ideal $In_2O_3$ TFTs. Current $In_2O_3$ TFTs show high-mobility (~100 $cm^2\ V^{-1}\ s^{-1}$). However, there is a major drawback; current $In_2O_3$ TFTs show large voltage shifts upon applying bias stress. Therefore, ideal $In_2O_3$ TFTs that show good reliability are in high demand. (c) Passivation strategy. Schematic illustration of a bottom-gate top-contact type $In_2O_3$ TFT with a surface passivation layer. ~50-nm-thick $HfO_2$, $Al_2O_3$, $Y_2O_3$, $Er_2O_3$, $Gd_2O_3$, $Yb_2O_3$, $Sm_2O_3$, and $Nd_2O_3$ films were tested as the surface passivation layer. C-rare earth oxides were mainly chosen because their crystal structures are the same as those of $In_2O_3$. $HfO_2$ and $Al_2O_3$ films were amorphous, whereas the other oxide films were polycrystalline.

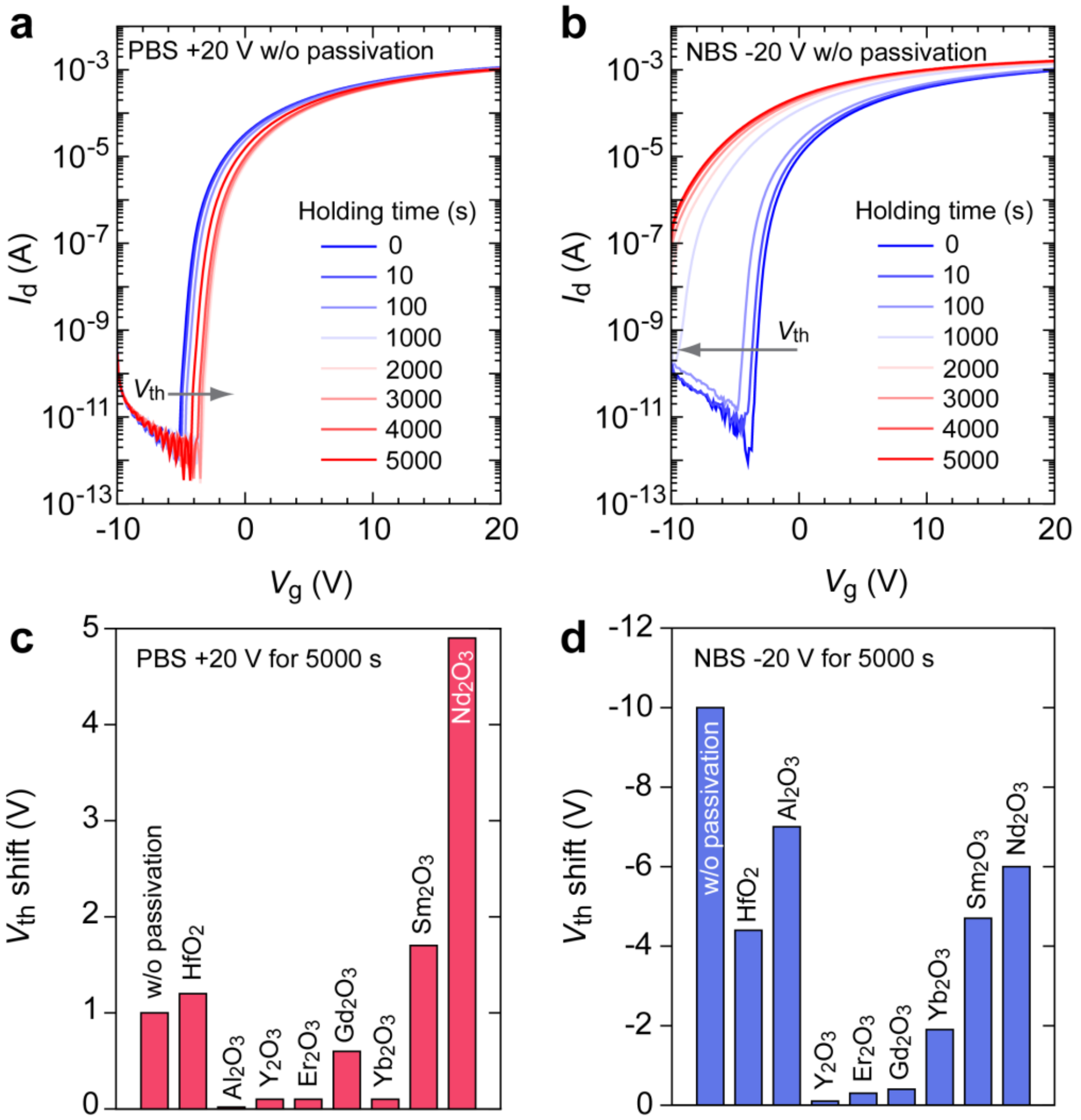

**Figure 2: Bias stress test results of the $In_2O_3$ TFTs with various passivation layers.** (a, b) Changes in transfer characteristics of the $In_2O_3$ TFTs without a passivation layer under (a) PBS (+20 V) and (b) NBS (−20 V). Threshold voltage ($V_{th}$) shifts of +1 V after 5000 s PBS and −10 V after 5000 s NBS were observed. This is the major drawback of $In_2O_3$ TFTs. (c, d) $V_{th}$ shifts of the $In_2O_3$ TFTs with various passivation layers after (c) PBS and (d) NBS for 5000 s extracted from **Figures S1−S8**. The $In_2O_3$ TFTs passivated with polycrystalline $Y_2O_3$ and $Er_2O_3$ films show negligible $V_{th}$ shift under both PBS and NBS. The $Gd_2O_3$ and $Yb_2O_3$ film passivation results in a small $V_{th}$ shift after the PBS and NBS. Overall, the $In_2O_3$ TFTs passivated with C-rare earth oxides show small $V_{th}$ shifts after PBS and NBS for 5000 s. Conversely, $In_2O_3$ TFTs passivated with $HfO_2$, $Al_2O_3$, $Sm_2O_3$, and $Nd_2O_3$ films still show large $V_{th}$ shifts after NBS, indicating that these films are not effective as the passivation layers for $In_2O_3$ TFTs.

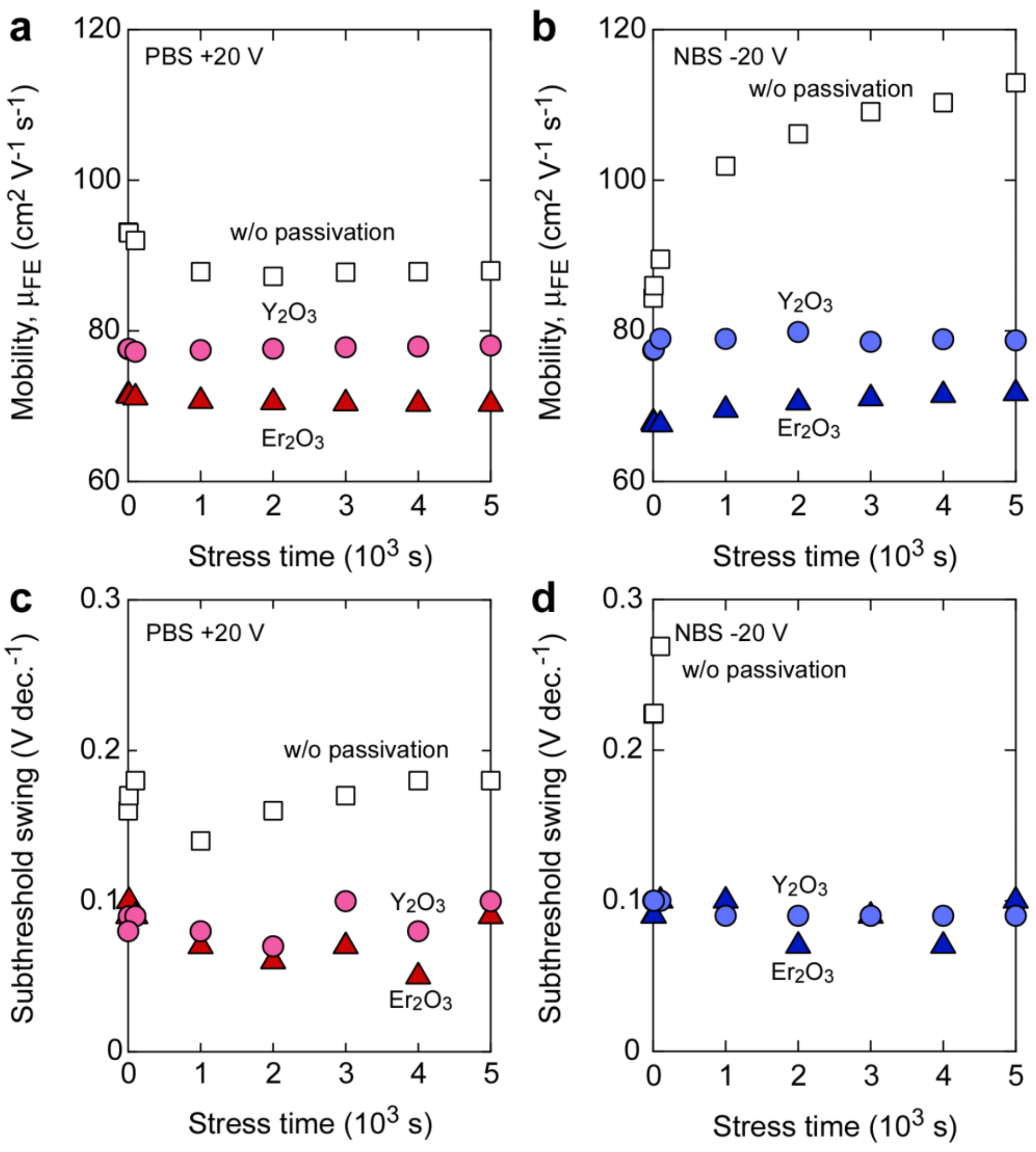

**Figure 3: Effects of the $Y_2O_3$ and $Er_2O_3$ films passivation on the transistor characteristics of the $In_2O_3$ TFTs.**

(a, b) Field effect mobility ($\mu_{FE}$). (c, d) Subthreshold swing. The $In_2O_3$ TFT without passivation exhibits unreliability; Both $\mu_{FE}$ and subthreshold swings change with the stress time. In contrast, the $In_2O_3$ TFTs passivated with $Y_2O_3$ and $Er_2O_3$ are highly reliable; the $\mu_{FE}$ and subthreshold swing are stable against the bias stress application. The $\mu_{FE}$ of the $Y_2O_3$ passivated TFT is ~78 $cm^2\ V^{-1}\ s^{-1}$ and that of the $Er_2O_3$ passivated TFT is ~70 $cm^2\ V^{-1}\ s^{-1}$. The subthreshold swing of both TFTs is ~0.1 V $decade^{-1}$.

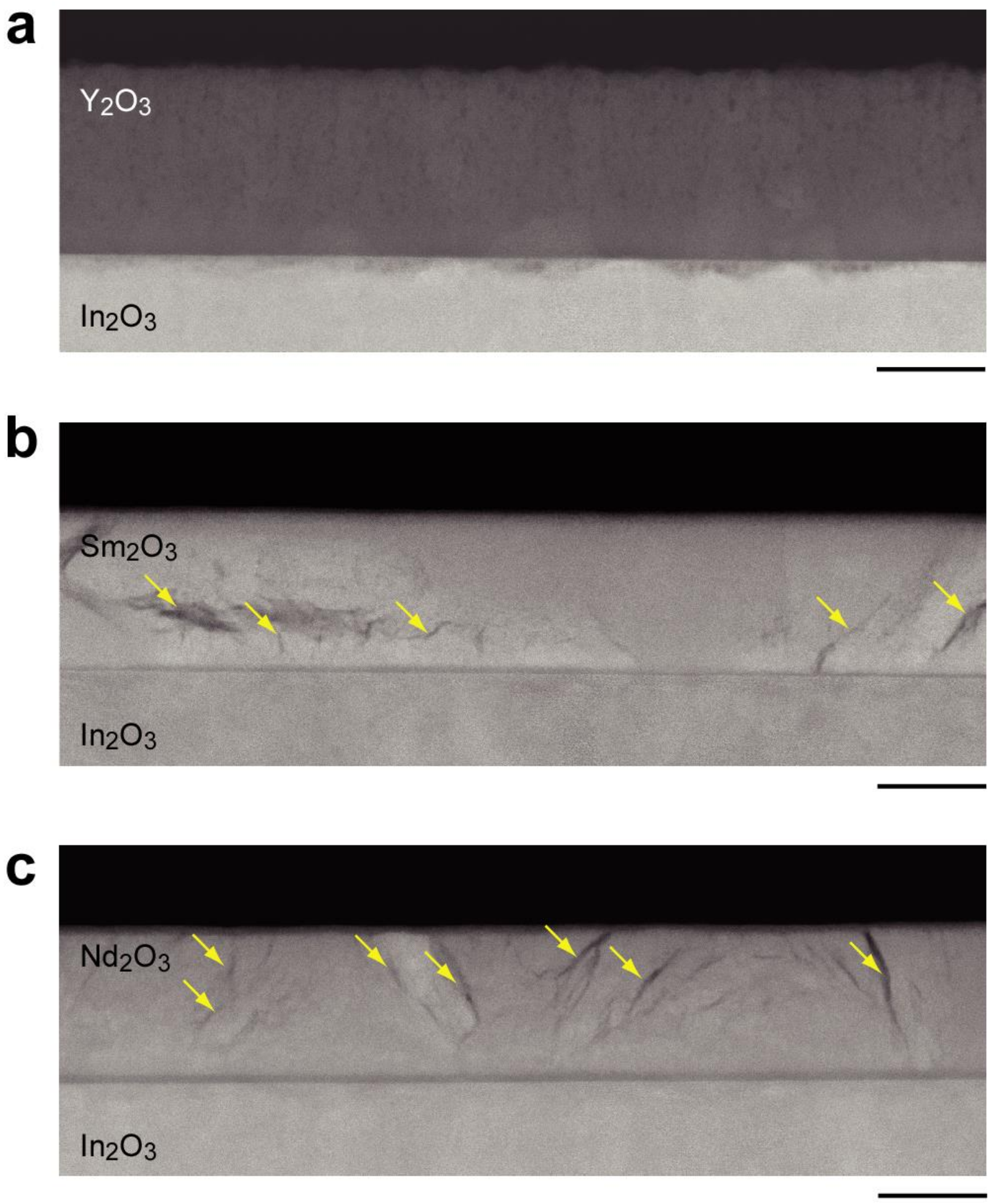

**Figure 4: Cross-sectional HAADF-STEM images of the *Ln*$_2$O$_3$ films on (111)-oriented In$_2$O$_3$ epitaxial films (10%-Sn-doped).**

(a) $Y_2O_3$ (C-rare earth), (b) $Sm_2O_3$ (B-rare earth), and (c) $Nd_2O_3$ (A-rare earth). The yellow arrows indicate the cracks. No cracks are observed in the $Y_2O_3$ film, while many cracks are observed in the $Sm_2O_3$ and $Nd_2O_3$ films. The incident e-beam direction is $\langle 1\bar{1}0 \rangle$ $In_2O_3$. The scale bars are 50 nm.

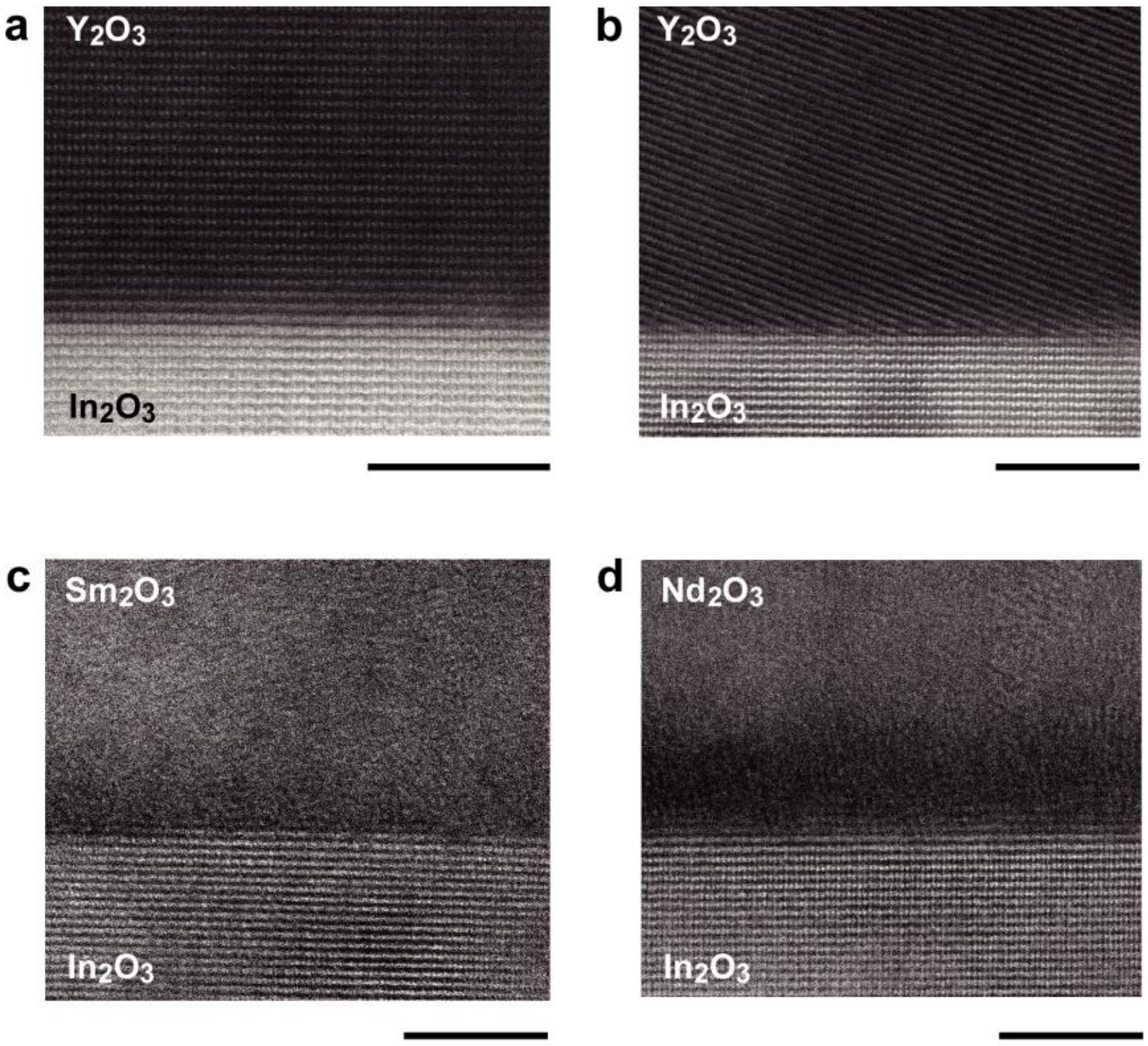

**Figure 5: Cross-sectional HAADF-STEM images (magnified) of the *Ln*$_2$O$_3$ films on (111)-oriented $In_2O_3$ epitaxial films (10%-Sn-doped).**
(a, b) $Y_2O_3$ (C-rare earth), (c) $Sm_2O_3$ (B-rare earth), and (d) $Nd_2O_3$ (A-rare earth). Lattice structures are observed in the case of $Y_2O_3$ film along with $In_2O_3$ lattice, indicating that heteroepitaxial growth occurs on the $In_2O_3$ crystal. In contrast, lattice structure is not observed in the case of $Sm_2O_3$ and $Nd_2O_3$ films. Further, thin amorphous layers are observed at the heterointerfaces of $Sm_2O_3/In_2O_3$ and $Nd_2O_3/In_2O_3$. The incident e-beam direction is $\langle 1\bar{1}0 \rangle$ $In_2O_3$. The scale bars are 5 nm.

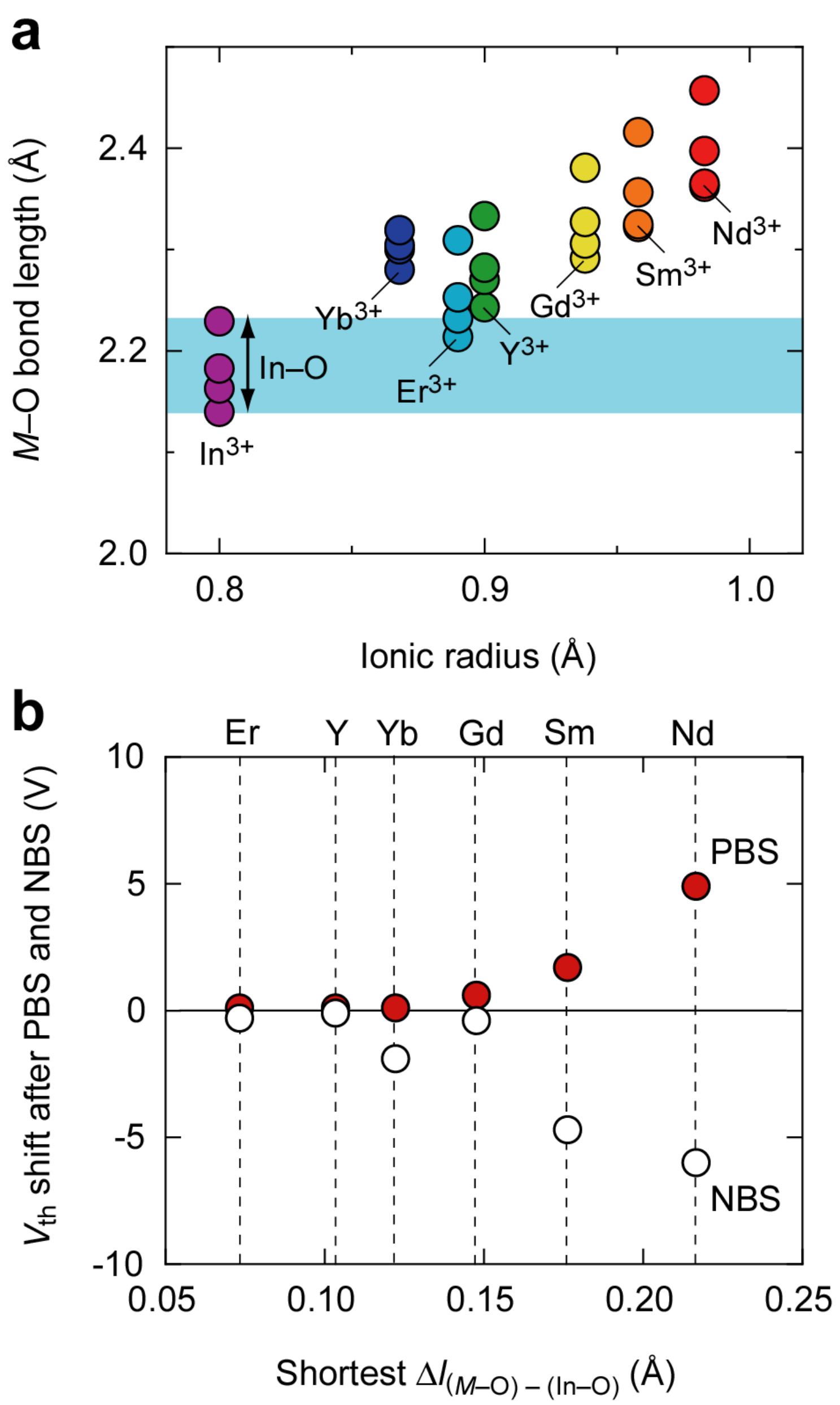

**Figure 6: Possible mechanism of suppression of bias stress shift.**
(a) Relationship between the ionic radius and metal−oxygen bond (*M*−O) length in various $Ln_2O_3$. The *M*−O length linearly increases with the expanding ionic radius of the metal. The light blue indicates the In−O length range. (b) $V_{th}$ shifts after PBS and NBS of the $In_2O_3$ TFTs as a function of the difference in *M*−O length from In−O length ($\Delta l_{(M\text{−O})-(\text{In−O})}$). The $\Delta l_{(M\text{−O})-(\text{In−O})}$ of the $Ln_2O_3$ were extracted from **Fig. 6a**. The $V_{th}$ shifts are significant for both PBS and NBS when the shortest $\Delta l_{(M\text{−O})-(\text{In−O})}$ of the passivation layer exceeds 0.15 Å ($Gd_2O_3$). In contrast, the $V_{th}$ shifts are negligible when the shortest $\Delta l_{(M\text{−O})-(\text{In−O})}$ is below 0.10 Å ($Y_2O_3$).

Supplementary Information

# Reliable operation in high-mobility indium oxide thin film transistors

Prashant R. Ghediya, Yusaku Magari*, Hikaru Sadahira, Takashi Endo, Mamoru Furuta, Yuqiao Zhang, Yasutaka Matsuo & Hiromichi Ohta*

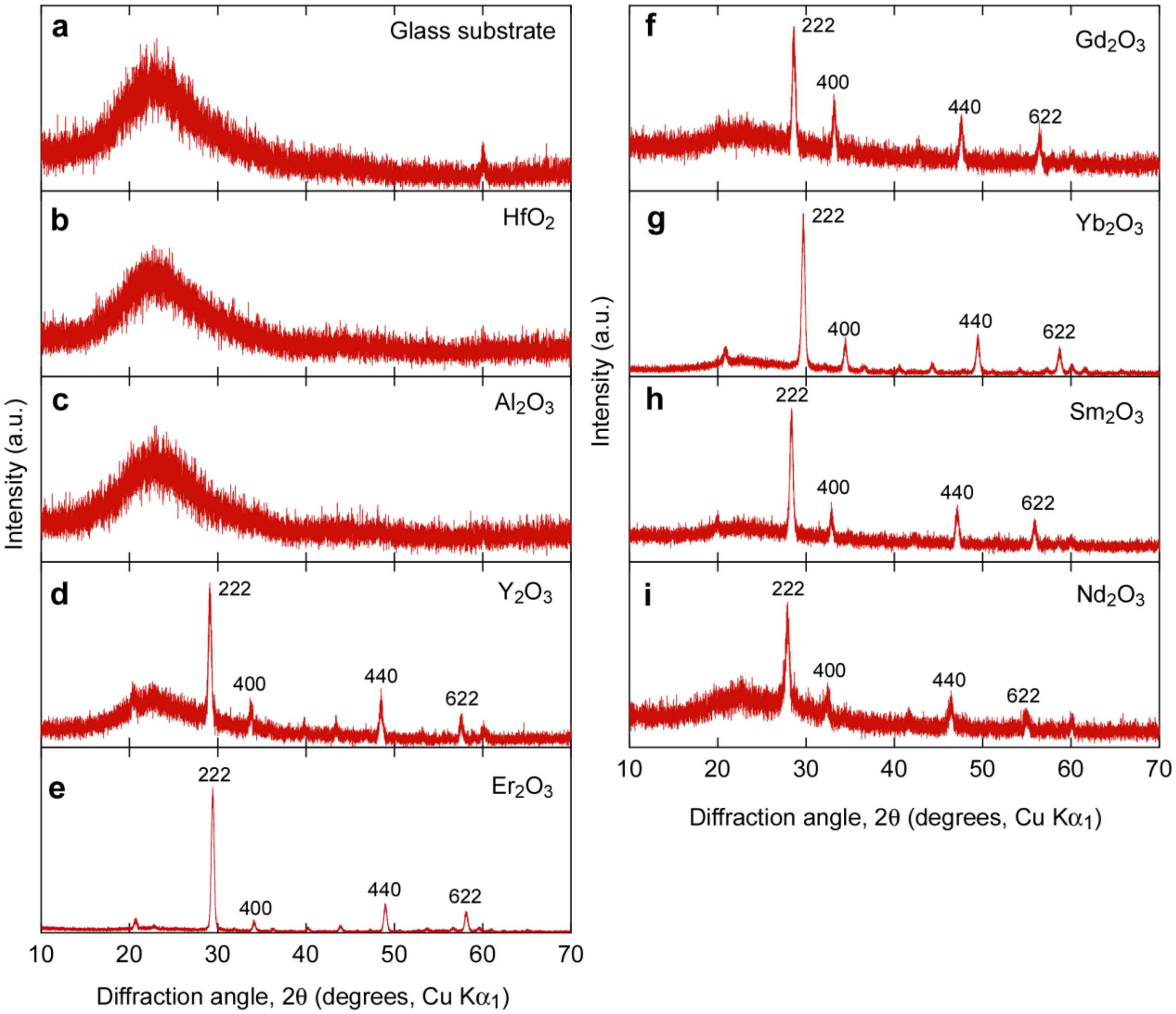

**Figure S1: XRD patterns of the passivation layers after thermal annealing at 370 °C.** The incident angle of the X-ray is 0.5°. (a) Glass substrate, (b) $HfO_2$, (c) $Al_2O_3$, (d) $Y_2O_3$, (e) $Er_2O_3$, (f) $Gd_2O_3$, (g) $Yb_2O_3$, (h) $Sm_2O_3$, and (i) $Nd_2O_3$. The passivation layers (100 nm) were deposited on alkali-free glass substrates and then thermally annealed at 370 °C in air. After the thermal annealing, $Ln_2O_3$ films were polycrystalline whereas $HfO_2$ and $Al_2O_3$ films remained amorphous.

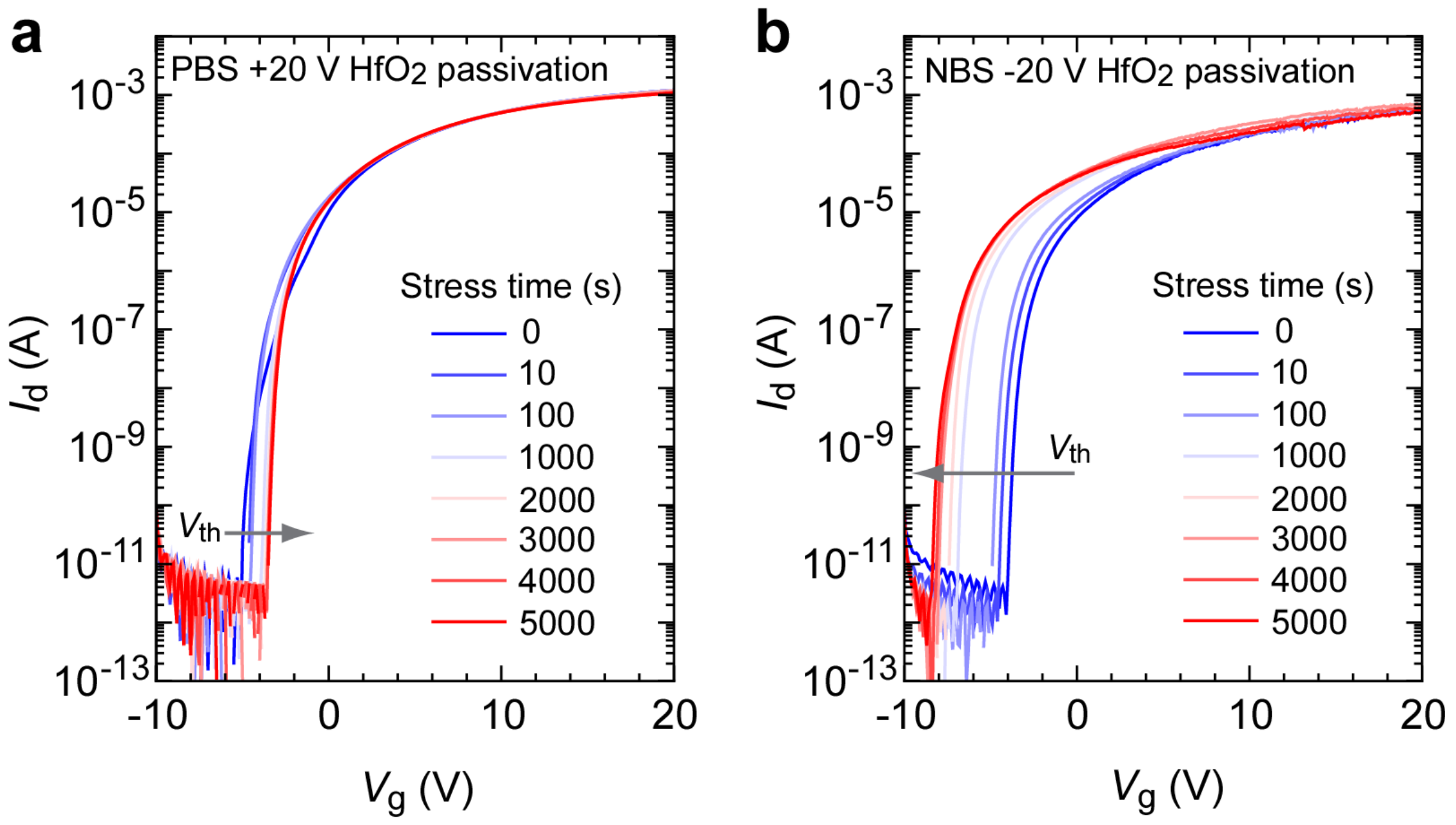

**Figure S2: Bias stress test results of the $In_2O_3$ TFT passivated with a $HfO_2$ film.** Changes in transfer characteristics under (a) PBS (+20 V) and (b) NBS (−20 V). Threshold voltage ($V_{th}$) shifts of +1.2 V after 5000 s PBS and −4.4 V after 5000 s NBS are observed, indicating that the $HfO_2$ film is not effective as the passivation layer for $In_2O_3$ TFTs.

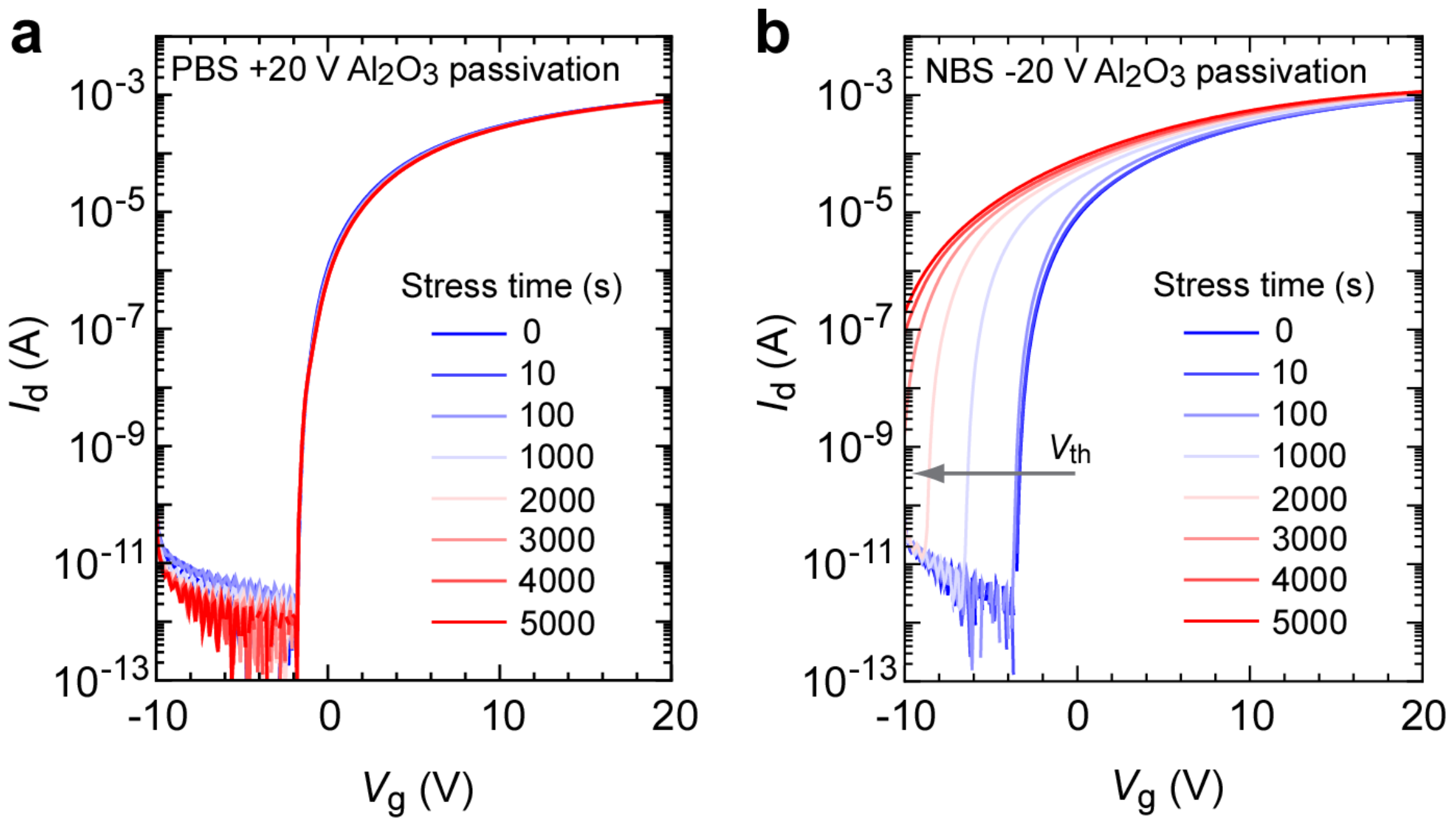

**Figure S3: Bias stress test results of the $In_2O_3$ TFT passivated with an $Al_2O_3$ film.** Changes in transfer characteristics under (a) PBS (+20 V) and (b) NBS (−20 V). Threshold voltage ($V_{th}$) shifts of +0.9 V after 5000 s PBS and −7 V after 5000 s NBS are observed, indicating that the $Al_2O_3$ film is not effective as the passivation layer for $In_2O_3$ TFTs.

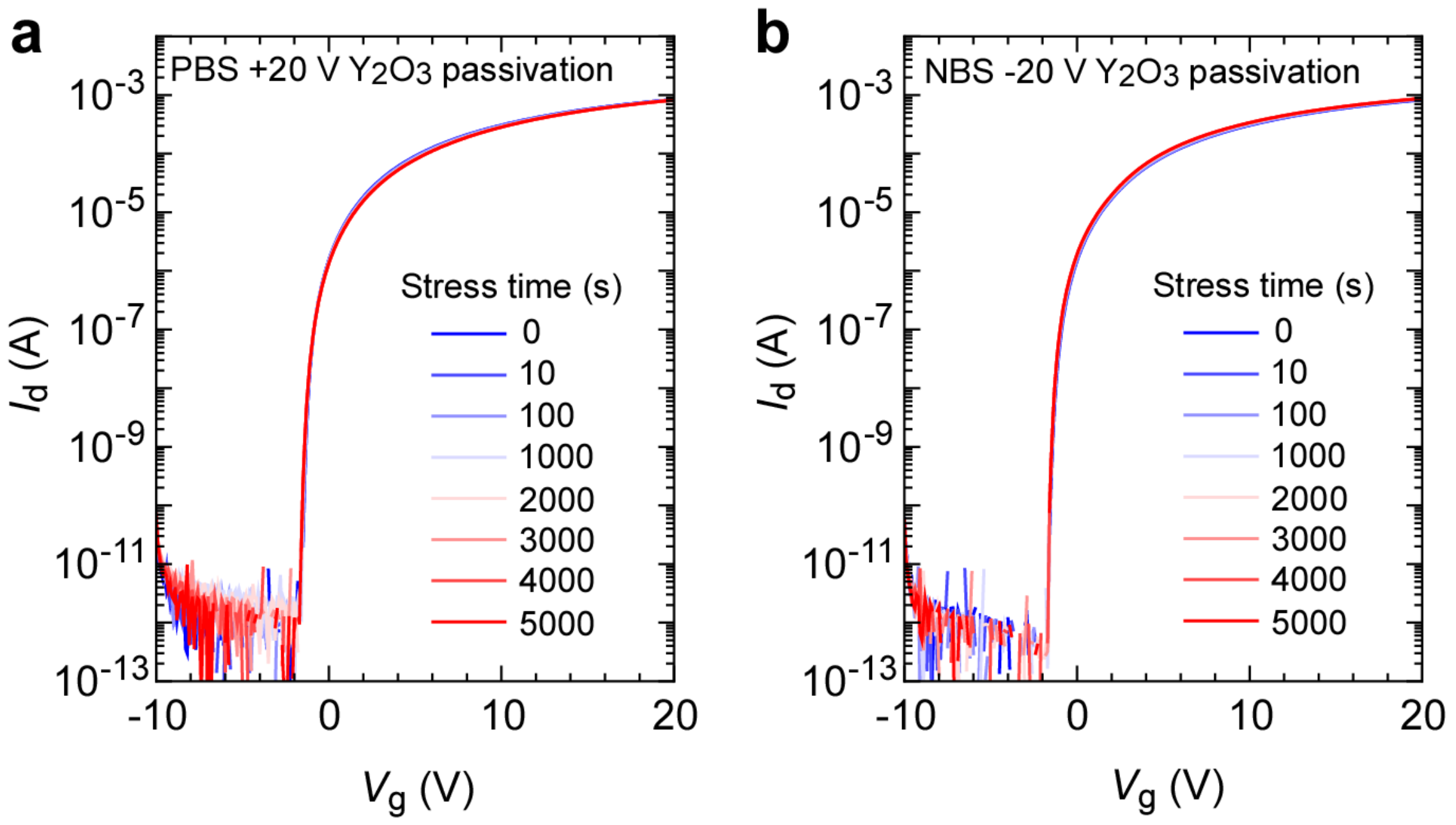

**Figure S4: Bias stress test results of the $In_2O_3$ TFT passivated with a $Y_2O_3$ film.** Changes in transfer characteristics under (a) PBS (+20 V) and (b) NBS (−20 V). The threshold voltage ($V_{th}$) shifts after 5000 s of PBS and 5000 s of NBS were negligible, indicating the high reliability of the $In_2O_3$ TFT passivated with the $Y_2O_3$ film.

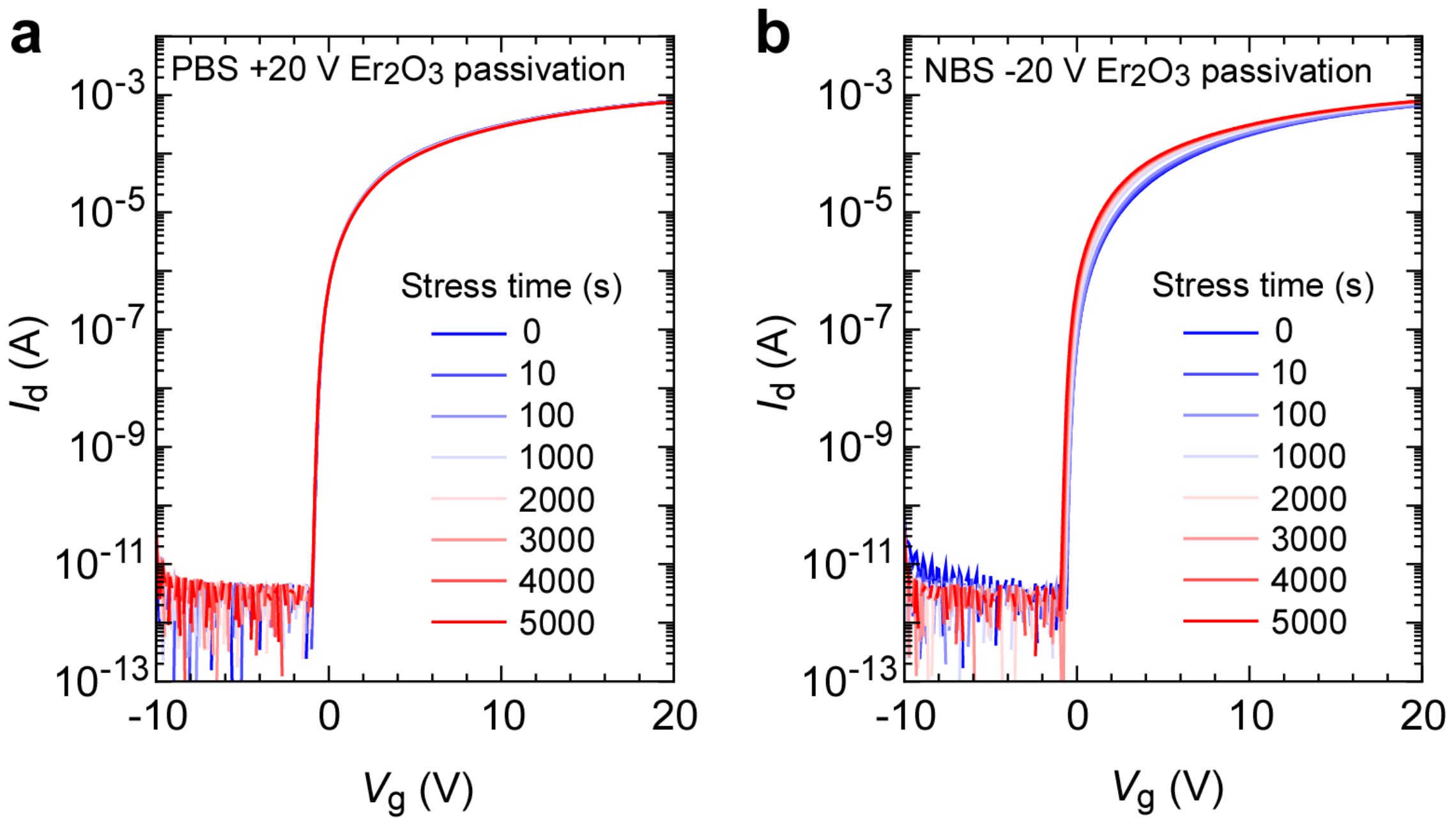

**Figure S5: Bias stress test results of the $In_2O_3$ TFT passivated with an $Er_2O_3$ film.** Changes in transfer characteristics under (a) PBS (+20 V) and (b) NBS (−20 V). The threshold voltage ($V_{th}$) shifts after 5000 s of PBS and 5000 s of NBS were negligible, indicating the high reliability of the $In_2O_3$ TFT passivated with the $Er_2O_3$ film.

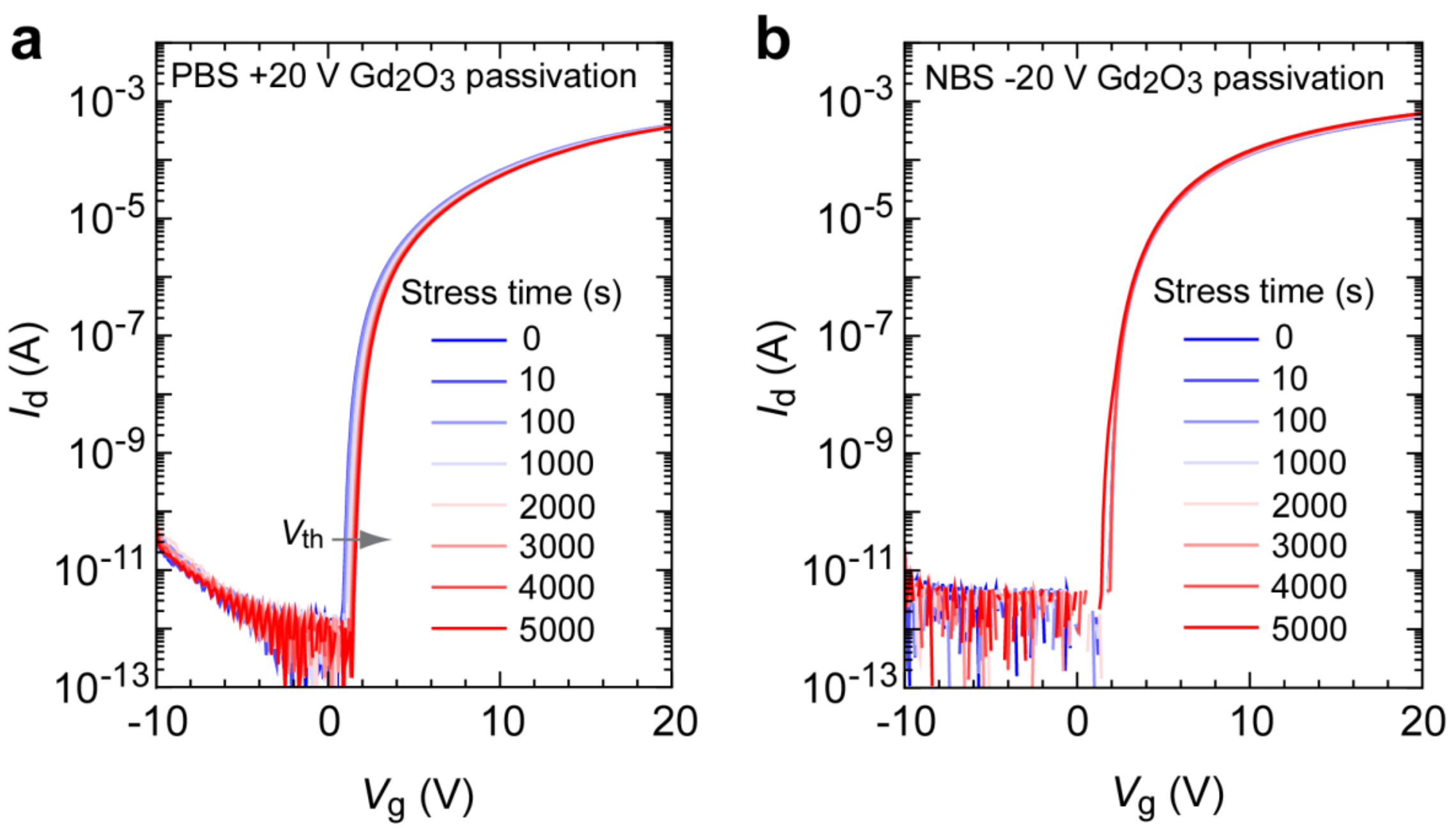

**Figure S6: Bias stress test results of the $In_2O_3$ TFT passivated with a $Gd_2O_3$ film.** Changes in transfer characteristics under (a) PBS (+20 V) and (b) NBS (−20 V). The $Gd_2O_3$ film passivation results in a +0.6 V $V_{th}$ shift after the PBS, whereas it results in a small $V_{th}$ shift (−0.4 V) after the NBS.

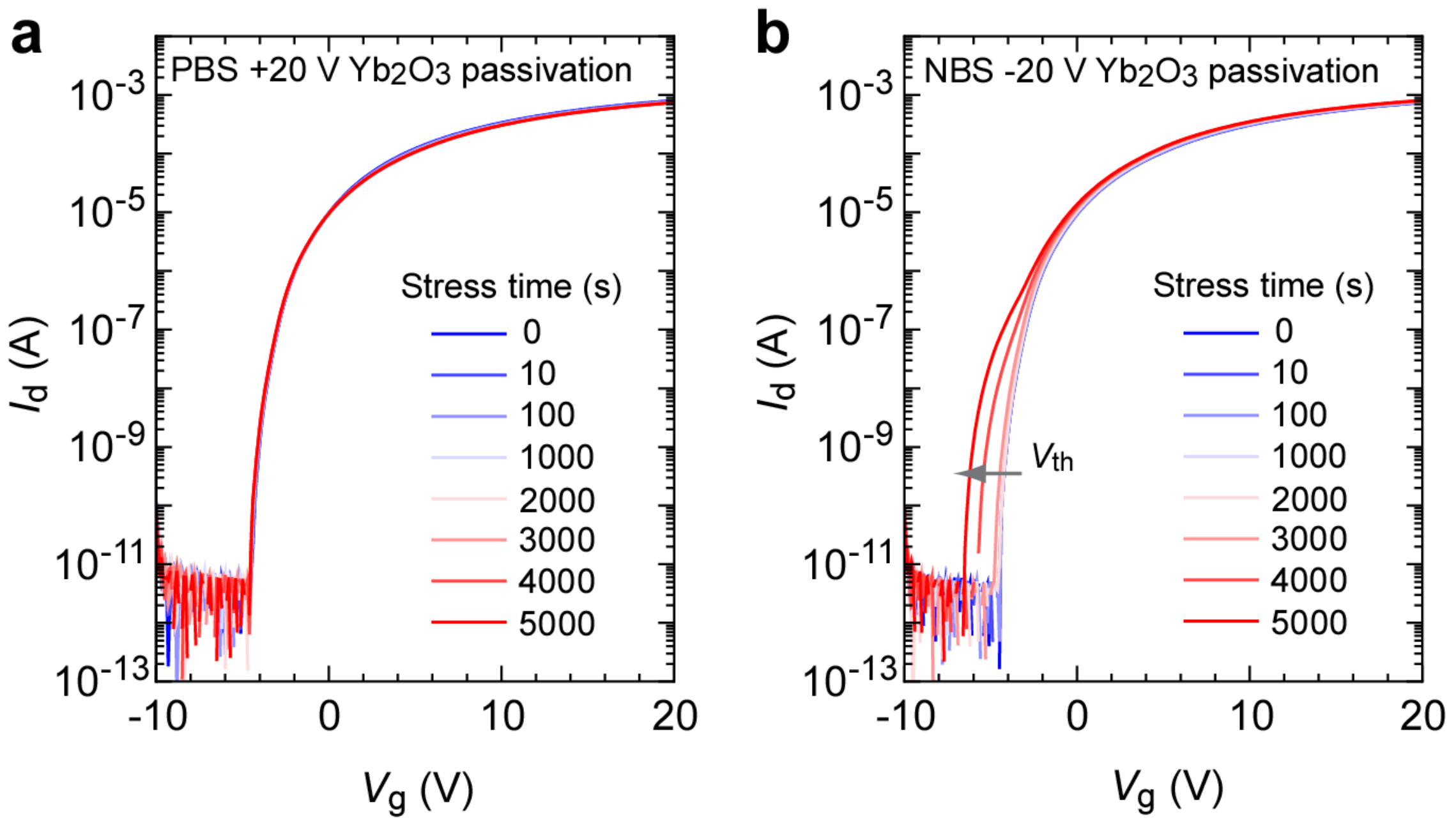

**Figure S7: Bias stress test results of the $In_2O_3$ TFT passivated with a $Yb_2O_3$ film.** Changes in transfer characteristics under (a) PBS (+20 V) and (b) NBS (−20 V). The $Yb_2O_3$ film passivation results in a negligible $V_{th}$ shift after the PBS, whereas it results in a rather large $V_{th}$ shift (−1.9 V) after the NBS.

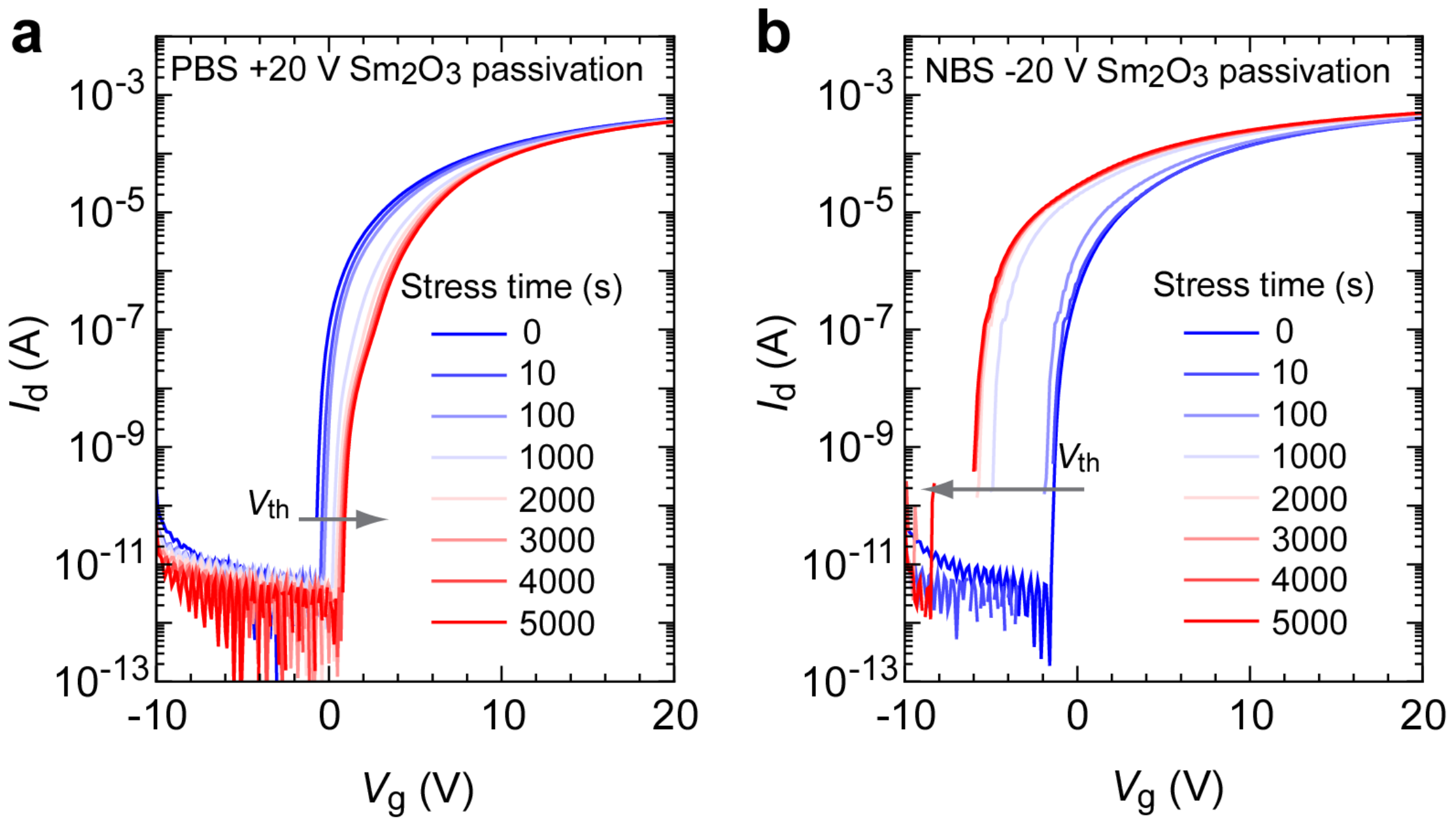

**Figure S8: Bias stress test results of the $In_2O_3$ TFT passivated with a $Sm_2O_3$ film.** Changes in transfer characteristics under (a) PBS (+20 V) and (b) NBS (−20 V). Threshold voltage ($V_{th}$) shifts of +1.7 V after 5000 s PBS and −4.7 V after 5000 s NBS are observed, indicating that the $Sm_2O_3$ film is not effective as the passivation layer for $In_2O_3$ TFTs.

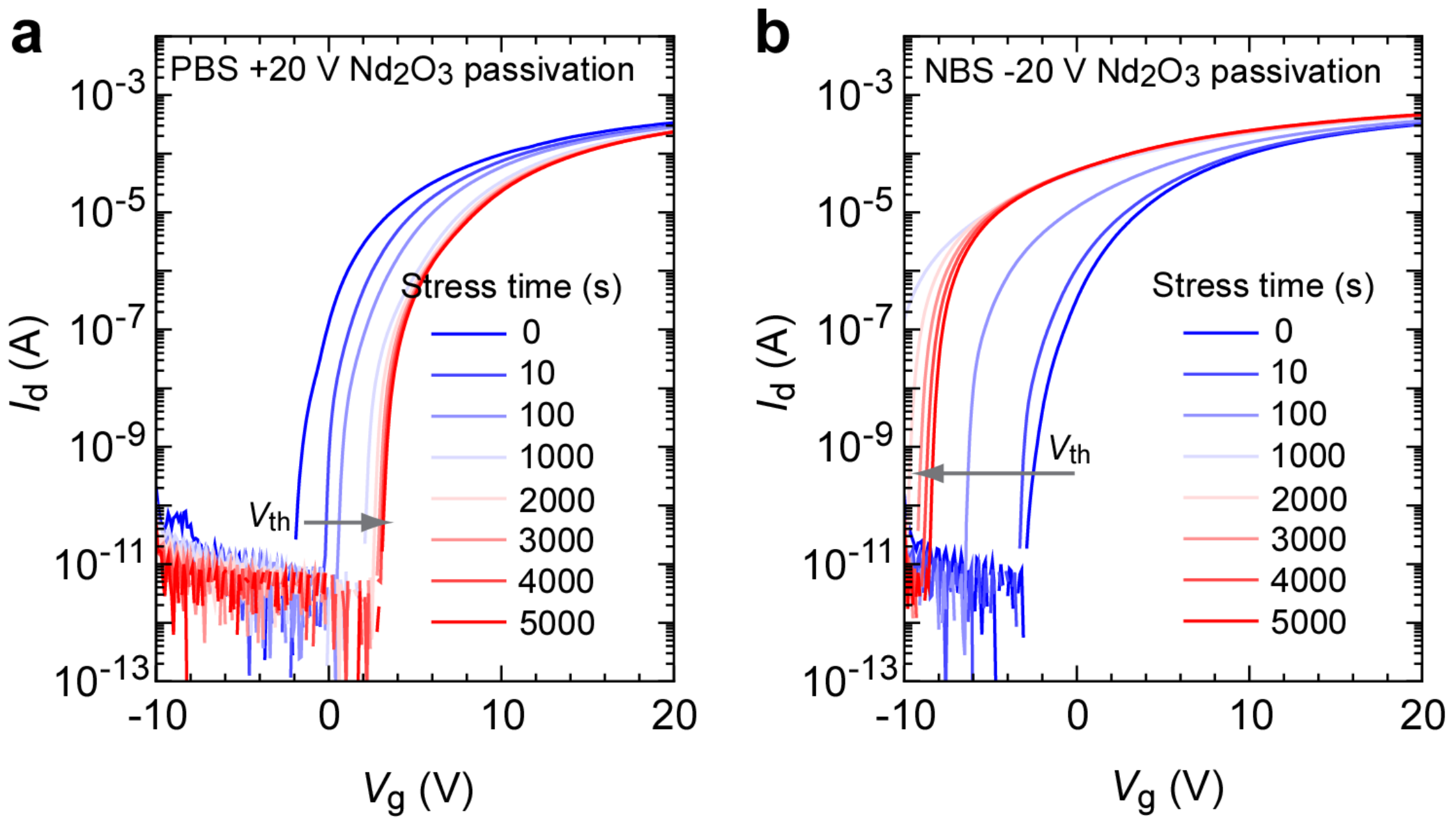

**Figure S9: Bias stress test results of the $In_2O_3$ TFT passivated with a $Nd_2O_3$ film.** Changes in transfer characteristics under (a) PBS (+20 V) and (b) NBS (−20 V). Threshold voltage ($V_{th}$) shifts of +4.9 V after 5000 s PBS and −6 V after 5000 s NBS are observed, indicating that the $Nd_2O_3$ film is not effective as the passivation layer for $In_2O_3$ TFTs.

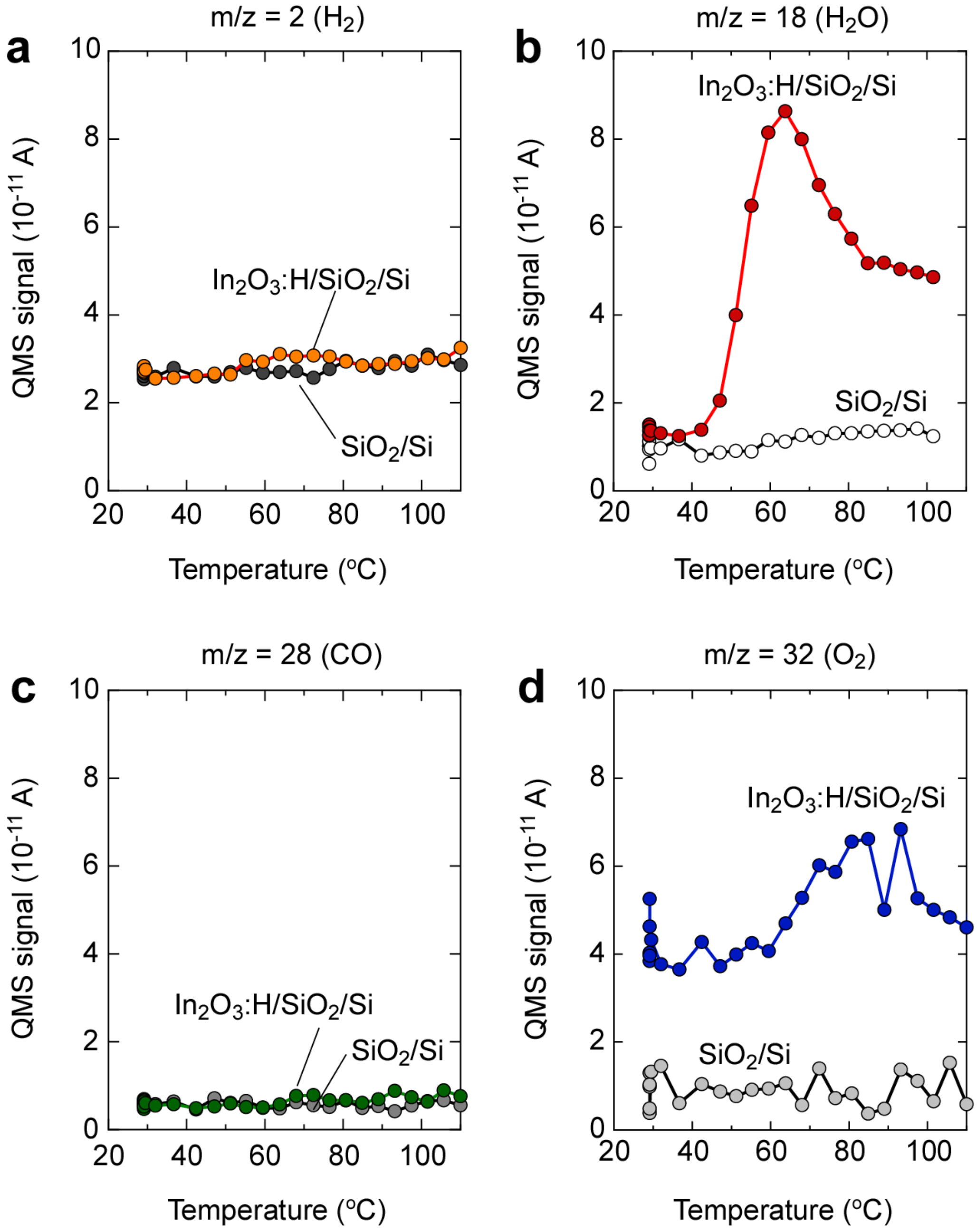

**Figure S10: Thermal desorption spectroscopy of $In_2O_3$ films.**
(a) m/z = 2 ($H_2$), (b) m/z = 18 ($H_2O$), (c) m/z = 28 (CO), and (d) m/z = 32 ($O_2$). A $SiO_2$/Si substrate was also tested. Large desorption peaks of $H_2O$ and $O_2$ are observed around 60 °C and 80 °C, respectively, whereas the desorption of $H_2$ and CO is negligibly small.

**Table S1: Comparison of reliability of $In_2O_3$ TFTs with various passivation layers.**

| Passivation | Structure | PBS (V) | NBS (V) | Judgment |
|---|---|---|---|---|
| Without | – | +1.0 | –10.0 | poor |
| $HfO_2$ | Amorphous | +1.2 | –4.4 | poor |
| $Al_2O_3$ | Amorphous | +0.02 | –7.0 | poor |
| $Y_2O_3$ | Polycrystalline (C-rare earth) | +0.1 | –0.1 | excellent |
| $Er_2O_3$ | Polycrystalline (C-rare earth) | +0.1 | –0.3 | excellent |
| $Gd_2O_3$ | Polycrystalline (C-rare earth) | +0.6 | –0.4 | excellent |
| $Yb_2O_3$ | Polycrystalline (C-rare earth) | +0.1 | –1.9 | fair |
| $Sm_2O_3$ | Polycrystalline (B-rare earth) | +1.7 | –4.7 | poor |
| $Nd_2O_3$ | Polycrystalline (A-rare earth) | +4.9 | –6.0 | poor |